\documentclass[letterpaper, 10 pt, conference]{ieeeconf}  

\IEEEoverridecommandlockouts                              

\usepackage{graphicx}

\usepackage{amsmath}

\usepackage{multirow}

\usepackage{footnote}
\usepackage{subfigure}
\usepackage{booktabs}

\title{\LARGE \bf
Clustering Human Trust Dynamics for Customized Real-time Prediction
}

\author{Jundi Liu$^{1}$, Kumar Akash$^{2}$, Teruhisa Misu$^{2}$ and Xingwei Wu$^{2}$  
\thanks{$^{1}$ Jundi Liu is with the Industrial \& Systems Engineering Department at the University of Washington, Seattle, WA 98105. This work was conducted during his internship at Honda Research Institute. 
        {\tt\small jundiliu@uw.edu}}%
\thanks{$^{2}$ Kumar Akash, Teruhisa Misu, and Xingwei Wu are with Honda Research Institute USA, Inc., San Jose, CA 95134, USA.
        {\tt\small \{kakash,tmisu,xingwei\_wu\}@honda-ri.com}}%
}

\begin{document}

\maketitle
\thispagestyle{empty}
\pagestyle{empty}

\begin{abstract}

Trust calibration is necessary to ensure appropriate user acceptance in advanced automation technologies. A significant challenge to achieve trust calibration is to quantitatively estimate human trust in real-time. Although multiple trust models exist, these models have limited predictive performance partly due to individual differences in trust dynamics. A personalized model for each person can address this issue, but it requires a significant amount of data for each user. We present a methodology to develop customized model by clustering humans based on their trust dynamics. The clustering-based method addresses the individual differences in trust dynamics while requiring significantly less data than personalized model. We show that our clustering-based customized models not only outperform the general model based on entire population, but also outperform simple demographic factor-based customized models. 
Specifically, we propose that two models based on ``confident'' and ``skeptical'' group of participants, respectively, can represent the trust behavior of the population. The ``confident'' participants, as compared to the ``skeptical'' participants, have higher initial trust levels, lose trust slower when they encounter low reliability operations, and have higher trust levels during trust-repair after the low reliability operations. In summary, clustering-based customized models improve trust prediction performance for further trust calibration considerations.
\end{abstract}

\section{Introduction}
Vehicle automation technologies have significant benefits for society, including dramatic decreases in car crashes, injuries and deaths, increased mobility, increased road efficiency, and better utilization of parking and lands \cite{mcgehee2016review, hamid2017autonomous}. Besides the advantages for society, they can also improve the driving experience and comfortableness of operations \cite{elmalaki2018sentio, kuderer2015learning}. Even though research has shown substantial benefits of vehicle automation technologies, their acceptance does not seem to keep up with the fast-growing market penetration. One of the most widely adopted methods to solve the acceptance issue is to calibrate trust in these technologies to the appropriate level since trust calibration is essential to accept and rely on vehicle automation \cite{niu2018anthropomorphizing}. Additionally, misuse of the system due to overtrust should be avoided \cite{de2014design}. Many studies have investigated the possible solutions to calibrate trust. These include paradigms that anticipate human behaviors---such as trust---and inform humans to make optimal choices \cite{drnec2016paradigm, metcalfe2017building, akash2020toward}. However, a primary challenge for such an approach is quantitatively predicting human trust in real-time.

Most current studies use questionnaires to measure self-reports of drivers' trust levels before, during, or after the interaction with the automated systems \cite{ruijten2018enhancing, chen2018planning, hoff2015trust}. However, it is challenging to obtain self-reports repeatedly without interrupting the task. As an alternative, recent works have developed dynamic models to capture human trust and estimate it in real-time \cite{muir1994trust, xu2015optimo, azevedo2020real, akash2020toward}. There are two approaches for developing such models: a general model for the whole population and a personalized model for each individual to account for individual differences \cite{wintersberger2016towards}. A general trust model ignores individual differences but can be trained using limited data. On the other hand, a completely personalized model designed for each person requires a significant amount of data for each new user. Such a personalized model may be applicable for some small-scale systems. However, for broad commercial applications, the amount of training data required is more than that can be collected in a short time period. Therefore, a tradeoff exists between limiting the amount of data needed for model training and improvement in model performance by personalization.

We address this tradeoff by using clustering methods to separate different trust dynamics across the sample population. We then develop customized trust models for each cluster of the population that account for broad individual differences in trust dynamics but allows model development with limited data. Specifically, we consider an interaction between a driver and a Society of Automotive Engineers (SAE) Level 2 driving automation and collect self-reports of trust throughout the interaction. We identify groups of users with critical differences in their trust dynamics using clustering based on trust evolution features. We demonstrate that the customized models based on these clusters significantly outperform the general model in predicting human trust as well as their take-over behavior. Additionally, although demographic factors have significant contributions to individual differences, we show that clustering based on trust dynamics-based features is more effective than simple demographic factor-based clustering for trust behavior prediction. Finally, we support the existence of the resulting clusters with literature from behavioral psychology. In summary, the contributions of this work are:
\begin{enumerate}
    \item a framework to cluster humans based on their trust behavior dynamics;
    \item identification of the ``confident'' and ``skeptical'' groups of users based on their trust in automation that is grounded in literature; and
    \item improvement in prediction performance of human trust and take-over behavior with limited data available using the clustering-based customized models.
\end{enumerate}
To the best of our knowledge, this is the first study that clusters users based on their trust dynamics, leading to an improved customized model for real-time trust prediction.

\section{Related Work}

With the fast emerging of vehicle automation technologies, there is skepticism rising in public concerns \cite{hengstler2016applied}. In a 2013 survey, 66\% of U.S. respondents indicated they were ``scared'' by the concept of automated driving, and more than half of respondents are skeptical of the reliability of the technology\cite{sommer2013continental}. The results show significant disagreements, that some people are more confident about the future of automated systems and others are still skeptical, have arisen in the attitudes toward the automated systems. From the human factors perspective, the vehicle automation needs to identify drivers' psychological characteristics and cognitive processes because those factors are reported to influence how drivers use these technologies \cite{dikmen2017trust}. Trust has emerged as a relevant focus in research since it provides a solid foundation for describing the relationship between humans and automation \cite{raats2019understanding, hengstler2016applied}. 

\subsection{Trust understanding and modeling}

Trust is shown to play a crucial role in understanding the acceptance of innovative technology \cite{luders2017innovating}. In fact, \cite{hoff2015trust} reported that trust determines the use or rejection of automation and willingness to rely on automation in certain situations.
Researchers have pointed out important aspects of trust in human-machine interaction. Muir et al. \cite{muir1994trust} showed that an individual's mental model has a strong relationship with the way he/she trust in the system. The research also emphasized that trust changes through experiences in association with the change of his/her mental model of the system. 

A well-accepted definition of trust in automation, proposed by Lee and See \cite{lee2004trust}, is ``the attitude that an agent will help achieve an individual's goals in a situation characterized by uncertainty and vulnerability''. They propose that the dynamic process rules trust and introduces context as a significant factor in trust development. Based on the definition of trust in \cite{lee2004trust}, Hoff and Bashir \cite{hoff2015trust} introduce three layers of trust: dispositional, situational, and learned trust. The dispositional trust is conceptualized based on early trust-related experiences that are typically affected by an individual’s demographics. Situational trust is context-dependent and is affected by situational information. The learned trust evolves with the experience with the system and differs by individual's mental model. This work is widely referenced in the later studies since they consider the trust evolution as a dynamic process \cite{schaefer2016meta, endsley2017here, casner2016challenges}. 

\subsection{Trust estimation and individual differences}

Most studies adopt trust-related questionnaires to obtain users' trust levels. Lee and Kolodge \cite{lee2020exploring} applied a topic model-based clustering method to comments about consumer attitudes towards vehicle automation. However, the questionnaires have a delayed effect and ignore the trust dynamics. While the factors discovered in the study provide guidance on feature extraction of trust dynamics, we aim at capturing individual differences in trust dynamics to achieve real-time prediction of user trust for trust calibration. 

Morra et al. \cite{morra2019building} proposed using a combination of questionnaire and galvanic skin response signal to quantify trust. The experiment was carried out using virtual reality as a human-machine interface to convey situational information, which is shown to improve trust in vehicle automation. Akash et al. \cite{akash2018classification} proposed a customized set of psychophysiological features for each individual to build a classifier-based trust-sensor model, which has investigated the trust estimation in real-time. Nevertheless, the psychophysiological measurements are intrusive and impractical in real-world implementation.
Several researchers have developed a variety of quantitative human trust models. These include regression models \cite{devries2003effects, muir1996trust}, time-series models \cite{moray2000adaptive,lee2004trust,hu2018computational}, and Markov models \cite{moe2008learning,elsalamouny2009hmmbased}. Recent work has demonstrated the use of a partially observable Markov decision process (POMDP) to model human trust dynamics to improve human-robot performance~\cite{akash2020human}. Researchers have also used a state-space model to capture human trust dynamics while interacting with a Level 3 driving automation based on automation performance, drivers' gaze, and drivers' non-driving related task performance \cite{azevedo2020real}. 

With the advancement of new methodologies, many new approaches have been proposed in recent years. Although there are studies that consider cultural differences that affect trust in automated systems \cite{hergeth2015effects}, previous studies on quantitative trust modeling often disregard the ``dispositional'' aspect. In this work, we demonstrate that individual differences in trust behavior (i.e., dispositional trust) can be captured by clustering the participants based on their trust dynamics (i.e., how they gain/lose the trust) to improve the performance of quantitative trust models.

\section{Online Study Design}

To model and cluster the dynamics of human trust, we collected human subject data using an online study where the participants interact with a Level 2 driving automation. The study used a simulated autonomous driving recording that was prerecorded using a physical driving simulator. The study was deployed on Amazon Mechanical Turk \cite{amazon2005amazon}, and the participants accessed the study online using their personal computers. During the study, the autonomous car drove through a series of ten intersections in an urban environment. The participants could press the spacebar key on their keyboard to indicate their intent to take over if they did not feel safe with the driving. Along with their take-over behavior, participants were also asked to provide self-reports of their trust as well as the reliability of the automation after each intersection during the study. Additionally, they completed a 12-question 7-point Likert scale pre-study and post-study trust questionnaire adapted from \cite{jian2000foundations}. 

Two within-subject factors were varied for the ten intersections: automation reliability and pedestrian presence. Automation reliability was defined only in terms of car’s stopping behavior at an intersection for consistency. It had two levels: low reliability, where the car aggressively decelerates very close to the stopline (deceleration starting at $<25$ meters), and high reliability, where the car smoothly decelerates toward the stopline (deceleration starting at $>60$ meters). Note that the reliability of the driving automation can be varied by other factors as well. Pedestrian presence also had two levels: either pedestrians are present or absent at the intersection. The presence of pedestrians can increase the perceived risk by the drivers. The risk was randomly varied across the ten intersections to avoid any ordering effects.

\begin{figure}[t]
  \centering
  \includegraphics[width=.56\linewidth]{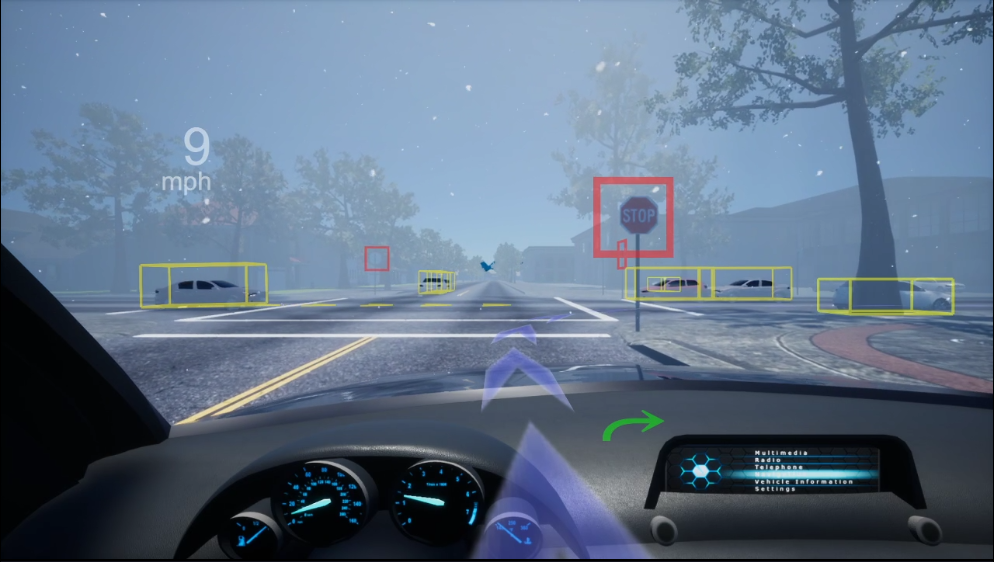}
  \caption{An example screenshot of the study scenario of low visibility and high automation transparency. The bounding boxes highlight the signs, cars and other moving objects in sight.}
  \label{fig:study_eg}
  \vspace{-5mm}
\end{figure}

Additionally, three factors that can affect human trust dynamics were varied between participants: scene visibility, overall reliability, and automation transparency. First, weather condition is shown to impact the trust levels significantly \cite{sheng2019case}. Scene visibility was varied by changing the weather of the environment and had two levels: high visibility where the scene was sunny and low visibility where the scene was foggy with snow. Second, situational characteristics such as automation reliability can also substantially impact trust dynamics \cite{noah2017trust}. In particular, users gain more trust in systems that are reliable. 
Overall reliability was designed to be affected by scene visibility such that the reliability of the automation reduces in low visibility scenes (as expected in real scenarios due to degraded scene perception). Specifically, the overall reliability had three levels: $100\%$ reliable, where none of the intersections were of low reliability and only occurred during high visibility scenes; $80\%$ reliable, where two intersections were of low reliability and occurred during both low and high visibility scenes; and $60\%$ reliable, where four intersections were of low reliability and only occurred during low visibility scenes. The low reliability intersections were randomly chosen across the ten intersections.
Third, studies have shown automation transparency also has a positive correlation with trust levels. With a higher level of transparency, the users have more access to the situational information leading to an increase in their trust \cite{mercado2016intelligent}. Studies have shown changing the level of automation transparency can calibrate trust to the appropriate levels \cite{akash2020human, yang2017evaluating}. In our study, during the high automation transparency, augmented reality (AR) cues are shown to the participant representing the driving automation’s perception of the scene. The AR cues can provide vehicle speed information, navigation information, and object detection and prediction that are present in the scene. In low transparency, the object detection and prediction are not presented. Fig. \ref{fig:study_eg} shows an example screenshot of the actual study scenario. Tab. \ref{tab:study_design} shows the resulting eight drive types and their corresponding characteristics. Tab. \ref{tab:event_config} shows the randomized distribution of reliability and risk in each drive type. Refer to the supplementary video for further demonstrations. 

\begin{table}[t]
\centering
\caption{Eight drive types in the online study. Overall reliability is the percentage of high reliability operations.}
\label{tab:study_design}
\vspace{-2mm}
\begin{tabular}{llll}
\toprule
Drv. Type & Overall Reliability & Visibility & Transparency \\ \midrule
A & 100\% & High & High \\
B & 80\% & High & High \\
C & 80\% & Low & High \\
D & 60\% & Low & High \\
E & 100\% & High & Low \\
F & 80\% & High & Low \\
G & 80\% & Low & Low \\
H & 60\% & Low & Low \\ \bottomrule
\end{tabular}
\vspace{-2mm}
\end{table}

\begin{table}[t]
\centering
\scriptsize
\caption{Event configuration in each intersection. The crosses denote the low reliability operation, and the Ps denote the presence of pedestrians in the specific intersection.}
\label{tab:event_config}
\vspace{-2mm}
\begin{tabular}{|c|c|c|c|c|c|c|c|c|c|c|}
\hline
Drv. Type & 1 & 2 & 3 & 4 & 5 & 6 & 7 & 8 & 9 & 10 \\ \hline
A &  & P &  &  & P &  & P & P & P &  \\ \hline
B &  & P &  & P & X P & P &  & X & P &  \\ \hline
C &  & P &  &  & X P & P & P & P & X &  \\ \hline
D &  &  & X & X P & X P &  & P &  & P & X P \\ \hline
E &  &  &  &  & P &  & P & P & P & P \\ \hline
F & P &  & P & P &  &  & P & P & X & X \\ \hline
G & P &  &  &  &  & X P & P & P &  & X P \\ \hline
H & P & P &  & X & P & X & P & X & X P &  \\ \hline
\end{tabular}
\vspace{-3mm}
\end{table}

\emph{Participants: }Two hundred thirty nine participants (121 males, 113 females, and 5 unknown) with ages between 19 and 77 years (mean: 39 years) from the United States participated in and completed the study online. They were recruited using Amazon Mechanical Turk, with the criteria that they must live in the US and have completed more than 1000 tasks with at least a 95\% approval rate. The compensation was \$2.25 for their participation, and each participant electronically provided their consent. The Institutional Review Board at Purdue University approved the study. Each participant completed a randomly selected drive type from Tab. \ref{tab:study_design}. Before the participants began the trial, they were given brief instructions about the study, and they completed a tutorial consisting of four intersections that helped familiarize them with the study interface. To ensure a uniform notion of trust across participants, they were explicitly informed about the definition adapted from \cite{lee2004trust} as follows: \\
\textit{``Trust is defined as your attitude that the self-driving car will help you achieve your goal of driving safely in a situation characterized by uncertainty and vulnerability.''}

Since the participants were not monitored during the study, we asked the participants to complete the study in fullscreen mode to avoid distractions. To avoid non-complying participants, we tracked the key-presses on the keyboard during the trial and removed the participants from the dataset who were suspected of exiting the fullscreen mode during the study. Furthermore, we removed the participants from the dataset who had missing survey data. As a result, 40 participants' data were removed from the dataset.

To summarize the online study design, we collected real-time trust levels of the users along with their take-over behavior while interacting with a Level 2 driving automation in a simulated environment. The study design considers three drive-level factors that can significantly impact the drivers' trust dynamics. Moreover, two event-level factors are considered within each drive type. The collected data will be used to identify the key characteristics of users trust dynamic, which will be the basis of clustering users based on their trust behavior.

\section{Methodology}


Using the intersection-by-intersection trust measurements obtained from the participants, we analyze the trust dynamics of each individual. We observe significant variations across individual trust behavior. Specifically, some participants start with lower initial trust levels as they may be skeptical about the automation. As they interact with a consistently high reliability automation, their trust levels increase gradually. However, if they encounter a low reliability operation, their trust levels drop drastically. As an extreme case, some participants' trust levels remain low throughout the study and do not increase significantly even after experiencing high reliability operation. On the contrary, some participants start with high initial trust levels and maintain stable and high trust levels throughout the study.

\subsection{Trust Evolution Decomposition}

We visualized the collected trust data to gain insights into the trust dynamics. Since there are drive types that do not have the low reliability operations or the low reliability operations happen at the end of the drive, we remove the participants from those drives to ensure all the participants for analysis have encountered the similar events in the study. As a result, we removed drive type A, E, and F from the samples, which left with 138 participants for further analysis. Based on the observations from the participants' data, we noticed the trust dynamics had three general phases across all the participants. The first phase is the initial \textit{trust-building} phase. In this phase, the participants start from their initial trust levels and gradually gain some trust as they experience high reliability operations. The initial trust levels depend on the personal characteristics of the participants. For example, people who have indicated prior good experience with advanced vehicle automation systems in the pre-study survey (scores are greater than 5 out of 7) are likely to have a higher initial trust level (average initial trust level is 88.6). The second phase is the \textit{error-awareness} phase, which occurs when a participant encounters the low reliability operation of the automation. After observing that the automation is not perfect, participants typically lose their trust in automation drastically. Finally, the third phase is the \textit{trust-repair} phase. This phase follows the error-awareness phase during the interaction, where the participants regain their trust in the automation as they experience consistent high reliability operations after low reliability ones. The trust increase during the trust-repair phase is typically lower than the initial trust-building phase since the participants realize that the automation may not be perfect and is prone to errors.

\begin{figure}[t]
\centering
\includegraphics[width=.9\linewidth]{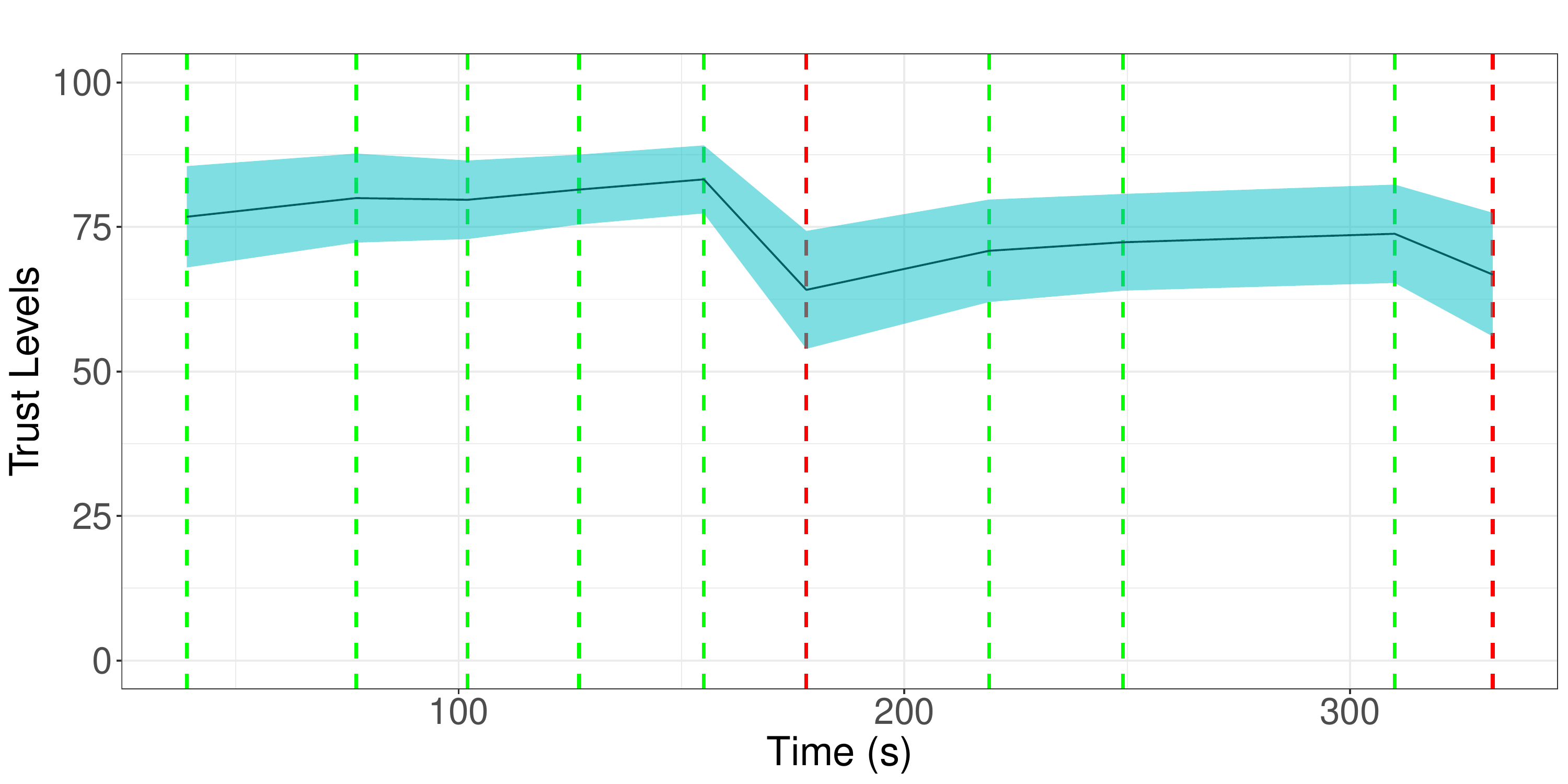}
\caption{Average trust dynamics for drive G. Each dashed vertical line denotes an intersection in the study. The green lines represent the high reliability operation, and the red lines represent the low reliability operation in that particular intersection. The black solid line denotes the average trust levels across all participants for drive G with the blue shaded area denoting the $95\%$ confidence intervals.}
\label{fig:trust_vis}
\vspace{-3mm}
\end{figure}

For example, Fig. \ref{fig:trust_vis} shows the average trust dynamics for drive G. In this drive, there are two low reliability operations at intersections 6 and 10. We observe that participants typically start with a relatively high average initial trust level ($\sim77$). This is consistent with findings that recent widespread use of automation has led to humans trusting a system when they have no experience with it \cite{merritt2008not}. During the initial trust-building phase (i.e., during consistent high reliability operation till intersection 5), the average trust level gradually increases for most of the participants, as apparent from smaller confidence intervals. At intersection 5, the trust level has the smallest confidence interval with the highest average value for this drive. Then in the two error-awareness phases (intersection 6 and 10), we observe a significant decrease in the trust levels and wider confidence intervals compared to the previous intersection, respectively. In this phase, participants notice the low reliability operations, which significantly reduces their trust level. Moreover, the first low reliability operation results in a much more significant effect in the decrease of trust level. Finally, the trust-repair phase (intersection 7 to intersection 9) shows the participants gradually regaining their trust after the low reliability operation if no further low reliability operation occurs. However, the rate at which the trust increases during the trust-repair phase is not as large as during the initial trust-building phase. Similar trends are also observed in other drive types.

In summary, we observed three main phases of the trust dynamics during the interaction with imperfect automation. However, the wide confidence intervals in Fig. \ref{fig:trust_vis} also suggest a considerable variation of the participants' trust levels across the three phases. In particular, the wide confidence intervals for the initial trust level show the participants have significant initial differences toward the automation. Although the confidence intervals converge during the trust-building phase, the confidence intervals are still wide during the error-awareness and trust-repair phases. These observations suggest that a general model may not be effective in capturing the individual differences in trust dynamics, and there is a need to develop customized models of trust.

\subsection{Trust Behavior Clustering}

To cluster the participants based on their trust dynamics, we first extract features characterizing the dynamics and behaviors of participants' trust throughout each interaction for each individual. Specifically, we extract the rate of change of trust to capture the trust dynamics for each of the three trust evolution phases. Furthermore, based on our preliminary analysis of trust, we observe that trust is strongly correlated to the presence of pedestrians at the intersection as well as participants' take-over response. Therefore, for the trust-building and trust-repair phase, we extract the average trust of participants at intersections where: 1) pedestrians are present, 2) pedestrians are absent, 3) the participants take-over 4) the participants do not take over. We additionally consider the initial trust level as a feature as it can capture the individual differences we observed in our analysis. This results in 12 features in total as listed in Tab. \ref{tab:features}. 

\begin{table}[t]
\centering
\caption{List of 12 extracted trust dynamics features.}
\label{tab:features}
\begin{tabular}{ll}
\toprule
Phase & Feature \\ \midrule
\multirow{6}{*}{Trust-building} & Initial trust level \\
 & Rate of change of trust\\
 & Average trust with pedestrian present \\
 & Average trust with pedestrian absent \\
 & Average trust with take-over \\
 & Average trust with no take-over \\ \midrule
\multirow{1}{*}{Error-awareness} & Rate of change of trust \\
 \midrule
\multirow{5}{*}{Trust-repair} & Rate of change of trust\\
 & Average trust with pedestrian present \\
 & Average trust with pedestrian absent \\
 & Average trust with take-over \\
 & Average trust with no take-over \\ \bottomrule
\end{tabular}\\
\vspace{-3mm}
\end{table}



Considering the limited sample size of unique participants\footnote{Although our data is not small to model trust behavior as each participant contribute to multiple trust samples, the number of unique participants (138) is limited for data-based clustering.}, we use simple Euclidean distance-based clustering method. To minimize the `curse of dimensionality' \cite{assent2012clustering} for the clustering algorithm, we use principal component analysis (PCA) to reduce the number of features used for clustering \cite{wold1987principal}. We apply PCA on the extracted features in Tab. \ref{tab:features} to reduce the dimension of data and provide insights on significant features that contribute most to the total variation. The explained variance ratios for the first three principal components (PC) are 44\%, 17\%, and 11\%, respectively \footnote{The variance ratios for the 4th and the subsequent components are 9\%, 6\%, etc.}. Thus, the first 3 PCs explain about 72\% of the total variance in the extracted features.

Finally, we use the K-means clustering method to find the groups of people with similar trust behaviors. K-means is one of the most widely used clustering methods, which iteratively computes each cluster's centroid and updates the assignment of each sample. While converging to stable assignments, K-means finds the clusters which minimize within-cluster variances. We chose the number of clusters as two since the silhouette analysis shows two clusters have the largest average silhouette coefficient (0.45) among all numbers of clusters varying from two to six and it ensures the best interpretability with significant statistical difference in all extracted features. With the identified clusters of participants, we will train customized models for each cluster to capture the individual differences of trust dynamics in each group. Furthermore, a close look into the clusters can provide insights into group-specific similarities.

\section{Results and Validation}

\begin{figure}
\centering
\subfigure[Initial trust level\label{fig_init_trust}]{\includegraphics[width=0.22\textwidth]{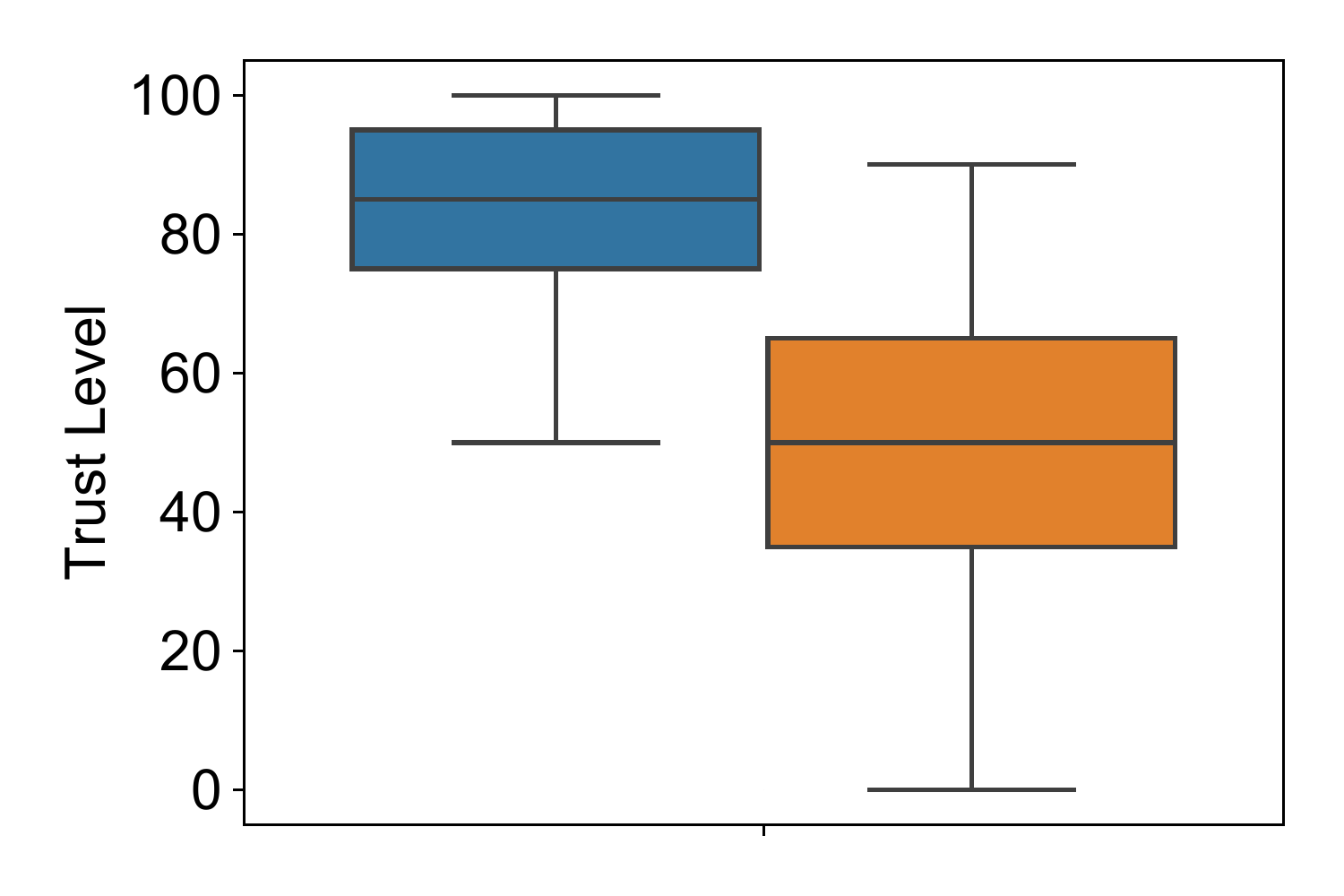}}
\subfigure[Average trust with pedestrian absent during trust-building\label{fig_init_nped_avg}]{\includegraphics[width=0.22\textwidth]{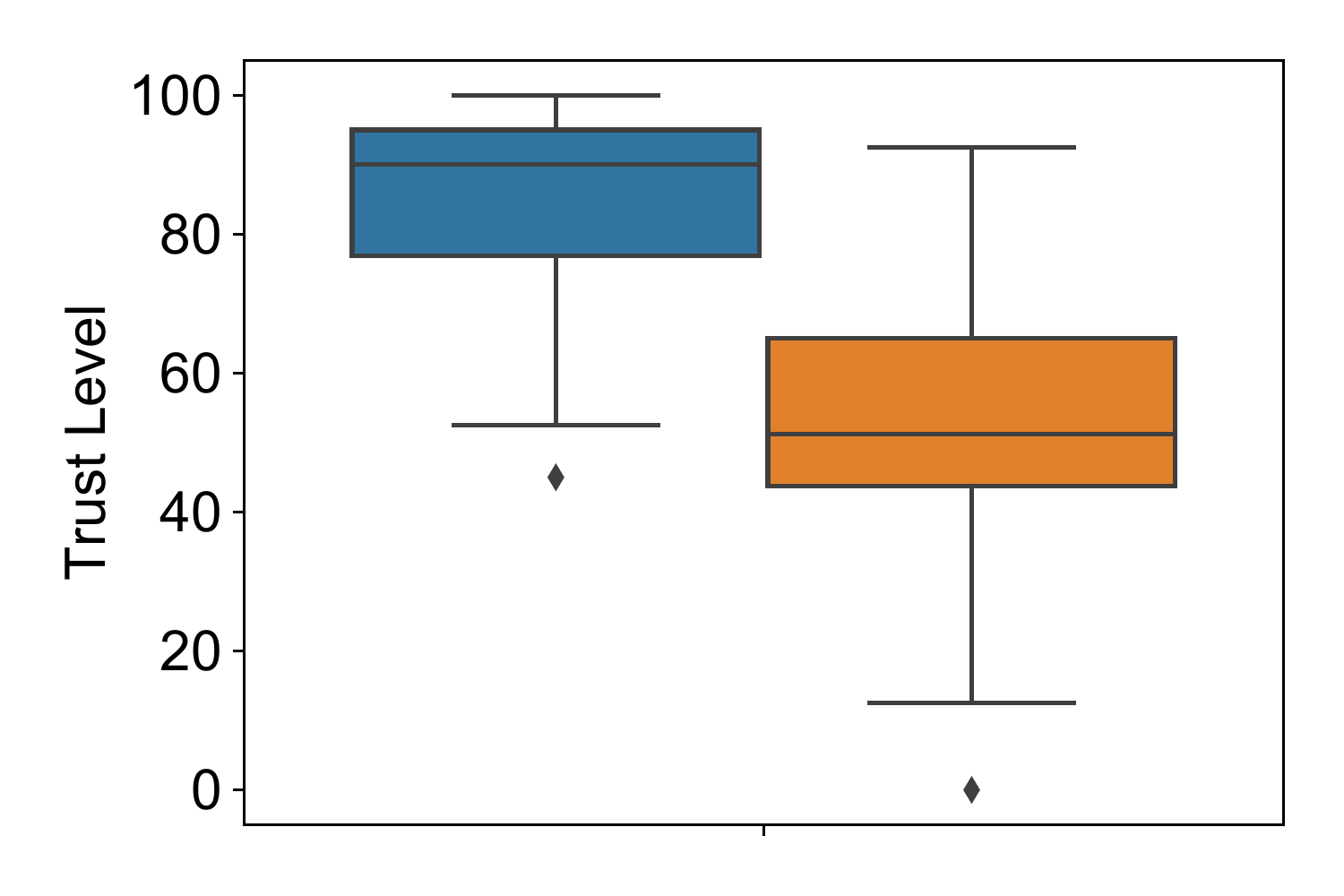}}
\subfigure[Rate of change of trust during error-awareness\label{fig_fall_spd}]{\includegraphics[width=0.22\textwidth]{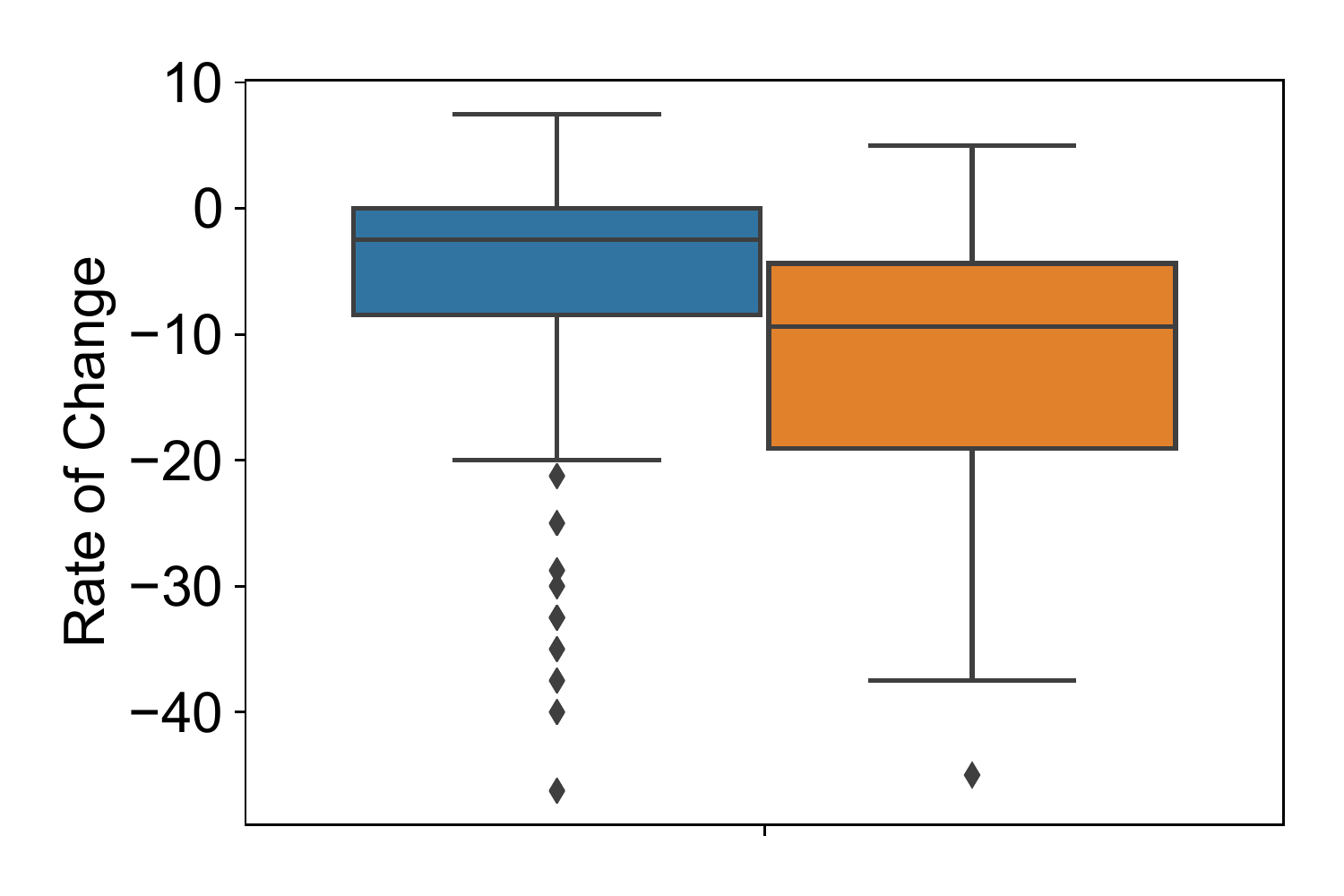}}
\subfigure[Average trust with no take-over during trust-repair\label{fig_regain_nbrk_avg}]{\includegraphics[width=0.22\textwidth]{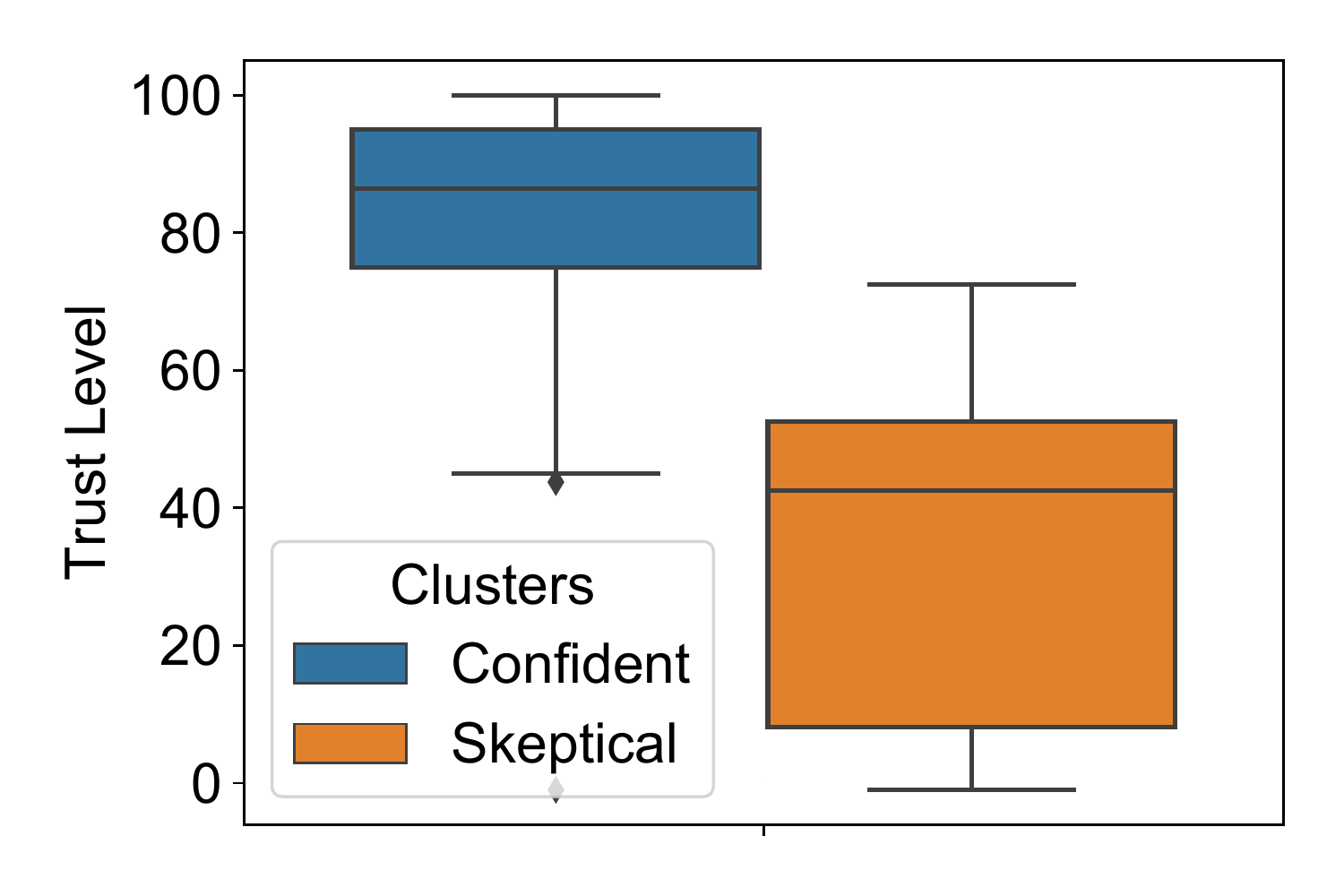}}
\caption{Boxplots of four representative features of trust dynamics. The two clusters show significantly difference from each other in all four features.}
\label{fig:box}
\vspace{-3mm}
\end{figure}

We applied the K-means clustering algorithm on the first three PCs to generate two clusters. The resulting two clusters have 36 and 102 participants, respectively. Fig. \ref{fig:box} shows the boxplots that demonstrate the variations in four representative features for the identified two clusters. For the initial trust (Fig. \ref{fig_init_trust}) and the average trust with pedestrian absence during trust-building (Fig. \ref{fig_init_nped_avg}), the two clusters show a statistically significant difference based on two sample t-test with a significant level of 0.05. Specifically, the cluster shown in orange has relatively lower initial trust than that shown in blue. Furthermore, the orange cluster also has a lower average trust with pedestrian absence during trust-building than the blue cluster. This shows that the participants comprising the orange cluster are more skeptical than those comprising the blue cluster. They tend to have a low initial trust toward the automation as well as a low trust level even during a low risk (due to the absence of pedestrians) and high reliability operations. Therefore, we name the orange cluster as the ``skeptical'' group due to their lack of trust. On the contrary, we call the blue cluster the ``confident'' group as they show high confidence on the driving automation. In the error-awareness phase, we observe from Fig. \ref{fig_fall_spd} that the ``skeptical'' group's trust levels are more volatile compared to the ``confident'' group and thereby drop faster after they encounter a low reliability operation. Finally, during the trust-repair phase, we consider the average trust level with no take-over (Fig. \ref{fig_regain_nbrk_avg}). This feature represents the trust level when the participants are comfortable with the automation system. We observe that the ``skeptical'' group has statistically lower trust levels than the ``confident'' group even when they do not take over. A two group t-test shows all 12 features are significantly different for the two clusters.


\begin{figure}[t]
\centering
\includegraphics[width=.9\linewidth]{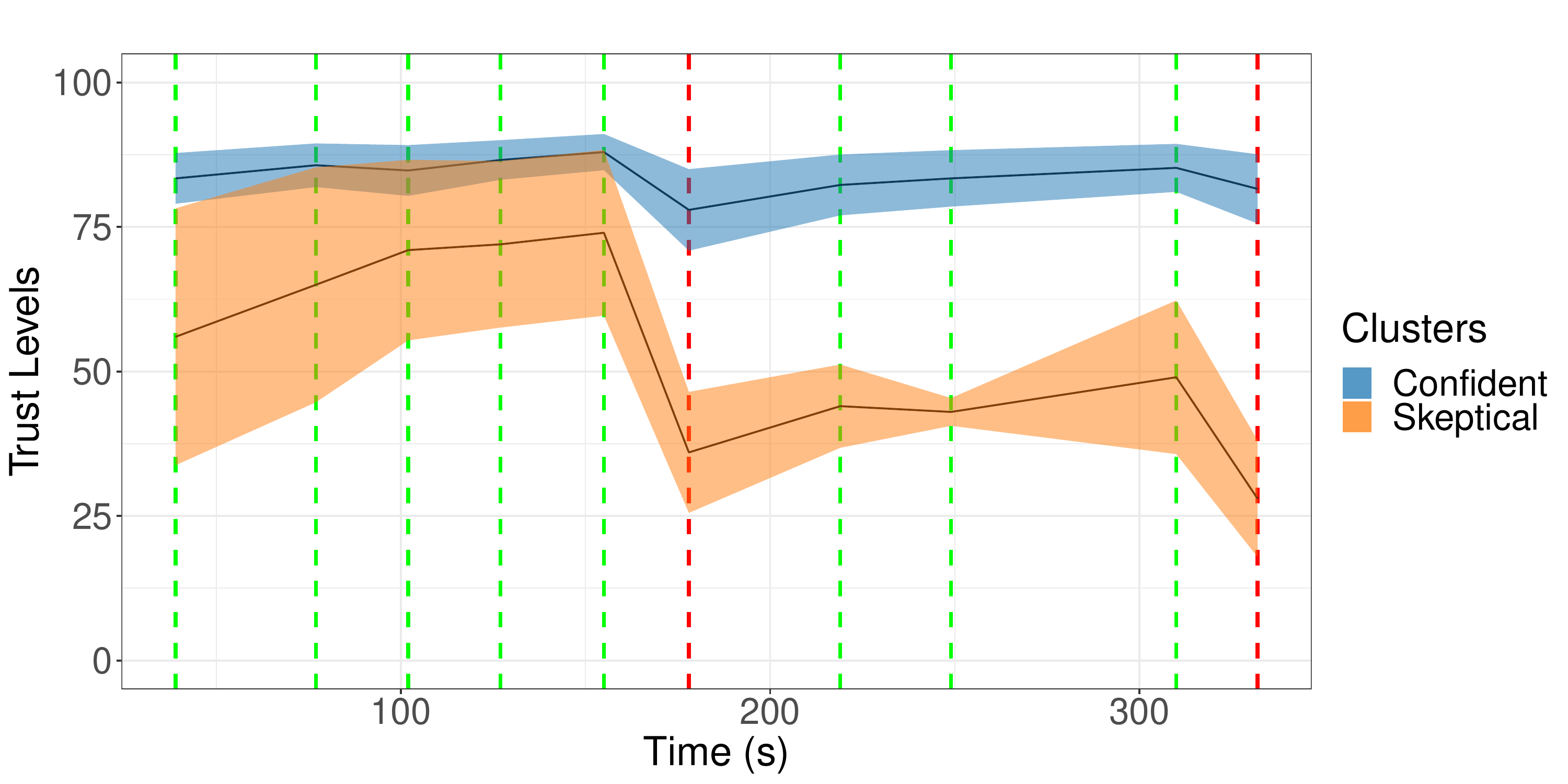}
\caption{Average trust level of the ``skeptical'' and the ``confident'' group participants at each intersection for drive G. The shaded area denotes the 95\% confidence intervals of the trust level.}
\label{fig:two_groups}
\vspace{-3mm}
\end{figure}

To better demonstrate the characteristics of the trust dynamics for the two groups, we compare the evolution of average trust for same drive type G as Fig. \ref{fig:trust_vis}in Fig. \ref{fig:two_groups}. The data for this drive type comprise of 22 ``confident'' group participants and 5 ``skeptical'' group participants. We see that the ``skeptical'' group start with a relatively lower initial trust and their trust level drops more quickly during a low reliability operation compared to the ``confident'' group.

While these clusters were identified using data-driven techniques, it is important to verify these behaviors with cognitive and behavioral psychology literature. Prior studies of both interpersonal human trust as well as human trust in automation note the existence of distinct trust behaviors among humans. 
Studies using the Rotter Interpersonal Trust Scale \cite{rotter1967new} have found two groups of trust behaviors, namely, ``high trusters'' and ``low trusters''. 
Furthermore, it was noted that characteristics of each individual varied in terms of willingness to trust a novel situation. Though the differences between humans' trust in automation was not explicitly analyzed, the other metrics established the two groups, such as high trusters being more willing to trust experimenters \cite{rotter1971generalized, rotter1980interpersonal}. 
It is likely that the ``confident'' and ``skeptical'' partially represent the high and low trusters, respectively.

\subsection{Customized Real-time Trust Prediction}

We show the improvement in trust prediction performance using customized models as compared to general model. The general model is trained using data for all participants and therefore, ignores the individual trust variations in trust dynamics. The customized models are separately trained using data from each group of participants identified by clustering, respectively. The resulting customized models leverage the distinct trust behaviors of each group that significantly improve the models performance. We consider two structures of trust prediction models for this comparison.

\subsubsection{Linear Regression (LR) Model: }
We first consider a simple linear regression model for predicting users' trust. We use the observations from the previous two intersections to predict the trust level at the current intersection. Specifically, we consider scene visibility, automation transparency, presence of pedestrians, automation reliability, participants trust level, and participants take-over behavior in the previous two intersections as an input to the model to predict the current trust level. Although this model may not be practical if self-reports of trust are unavailable, the models provides a simple baseline to validate the clustering performance.

\subsubsection{State Space (SS) Model with Kalman Filter: }
A classical approach to model a dynamical system is by using a linear time-invariant state-space (SS) model. Linear SS model has been used to capture human trust dynamics while interacting with a Level 3 driving automation based on automation performance, drivers' gaze, and drivers' non-driving related task performance \cite{azevedo2020real}. Considering trust as a continuous state of the system, they use Kalman filter to estimate trust during the interaction. Since our output of take-over intent is a binary variable, we adapt the linear state space model in \cite{azevedo2020real} with a sigmoid output function that maps the state of trust to the output of take-over. Thereby, we consider scene visibility, automation transparency, pedestrian presence, and automation reliability as inputs of the SS model; the state is a continuous variable of trust; and the output is the take-over intent. The model parameters are estimated using linear mixed-effect model with participants as a random effect as described in \cite{azevedo2020real}. Note that the self-reported trust is used as the measurement for the continuous trust state and is only used for model training. To evaluate the prediction performance of trust and take-over intent, we use an extended Kalman filter (EKF) that accommodates the nonlinear sigmoid output function to update the state estimate of trust after each output is observed, which is then used to predict the next take-over intent based on current inputs. Therefore, the EKF provides a real-time estimate of trust and take-over without the need for self-reports of trust during real-time prediction. The model is characterized by \eqref{equa:ekf}.
\begin{align}
\begin{split}
     T_{k+1}&=\mathbf{A}T_k+\mathbf{B}\begin{bmatrix}
           v_k &
           t_k &
           p_k &
           f_k
         \end{bmatrix}^T\\
     b_k &= \text{Sig}(\mathbf{C}T_k + \mathbf{C}_b)
    \label{equa:ekf}
\end{split}
\end{align}
Here $\text{Sig}(x)= (1+e^{-x})^{-1}$, 
$T_k$ is the trust level at the $k$th event, $v$ is the scene visibility, $t$ is the automation transparency, $p$ is the presence of pedestrians, $f$ is the automation reliability, $b$ is the take-over behavior of the participant, and $\mathbf{A}, \mathbf{B}, \mathbf{C}, \text{and}, \mathbf{C}_b$ are linear parameters.

\begin{table*}[t]
\centering
\caption{MSE and F1 scores for general model and customized models using different clustering criteria. Lower MSE and higher F1 score indicates better model performance.}
\label{tab:mse}
\begin{tabular}{l l c c c c }
\toprule
\multirow{2}{*}{Clustering Criteria} & \multirow{2}{*}{Cluster} & \multicolumn{1}{l}{\multirow{2}{*}{Number of participants}} & \multicolumn{2}{c}{MSE for trust} & F1 scores for take-over \\ \cmidrule{4-6} 
 &  & \multicolumn{1}{l}{} & LR model & SS model & SS model \\ \midrule
- & General model & 138 & 0.602 & 0.518 & 0.423 \\ \midrule
\multirow{2}{*}{\begin{tabular}[c]{@{}l@{}}Trust dynamics\\ (our method)\end{tabular}} & ``Confident'' & 102 & 0.442 & 0.381 & 0.468 \\ 
 & ``Skeptical'' & 36 & 0.384 & 0.289 & 0.650 \\ \midrule
\multirow{2}{*}{Age} & ``At least 40'' & 53 & 0.587 & 0.586 & 0.520 \\ 
 & ``Less than 40'' & 85 & 0.592 & 0.431 & 0.327 \\ \midrule
\multirow{2}{*}{Gender} & ``Male'' & 65 & 0.594 & 0.488 & 0.456 \\ 
 & ``Female'' & 71 & 0.526 & 0.549 & 0.396 \\ \midrule
\multirow{2}{*}{Driving Style} & ``Aggressive'' & 61 & 0.543 & 0.571 & 0.347 \\ 
 & ``Conservative'' & 77 & 0.526 & 0.493 & 0.460 \\ \bottomrule
\end{tabular}
\vspace{-3mm}
\end{table*}

Additionally, we compare the performance of our proposed trust dynamics-based clustering with demographic information based clustering. Specifically, we consider three baselines: 1) age with threshold as the mean age (40 years) of the participants sample; 2) gender (male or female); and 3) self-reported driving style (aggressive or conservative). We perform a 5-fold cross validation (CV) with uniform distribution of each drive type in the training and validation sets for both the model structures to obtain the validation performances. We calculate the validation mean squared errors (MSE) for trust prediction using both models and the F1 scores for take-over intent prediction using the SS model for the general model as well as for the clusters using each of the clustering criteria. The result is shown in Tab. \ref{tab:mse}. 


We see that for both the LR and SS models, the customized models based on the trust dynamics-based clustering performs better than the general model. That is, the trust dynamics-based customized models have lower MSE for trust as well as higher F1 score for take-over intent prediction. Therefore, the customized models successfully improves the trust prediction by considering the individual differences across the population. However, we do not see such significant improvement for age-based, gender-based, or driving style-based customized models. To further quantify the improvement in the prediction performance using the customized model, we calculate the percentage increase in the average metrics (MSE and F1 score) for each clustering criteria as compared to the general model. The average metrics for each clustering criteria is calculated as the mean of the metrics across the clusters weighted by the number of participants in each cluster. The resulting improvements in the prediction performance is shown in Tab. \ref{tab:change}. We observe that the trust-dynamics based customized models not only improves the prediction performance, it significantly outperforms simple demographic factor based clustering. This shows that these demographic factors alone may not be strong contributors to the variations in human trust behavior. 

\begin{table}[t]
\centering
\caption{Percentage increase in the prediction performance using the customized model as compared to the general model}
\label{tab:change}
\begin{tabular}{lccc}
\toprule
\multirow{2}{*}{Clustering Criteria} & \multicolumn{2}{c}{\begin{tabular}[c]{@{}c@{}}MSE for trust\end{tabular}} & \begin{tabular}[c]{@{}c@{}}F1 score for take-over\end{tabular} \\ \cmidrule{2-4} 
 & LR model & SS model & SS model \\ \midrule
\begin{tabular}[c]{@{}l@{}}Trust dynamics\end{tabular} & 29.1\% & 31.1\% & 21.7\% \\ 
Age & 2.0\% & 5.2\% & -5.2\% \\ 
Gender & 7.1\% & -0.3\% & 0.5\% \\ 
Driving style & 11.3\% & -1.7\% & -3.1\% \\ \bottomrule
\end{tabular}
\vspace{-5mm}
\end{table}

In summary, we demonstrate that trust dynamics-based clustering can allow to develop improved trust model needed for trust calibration paradigms. In practice, a small interaction data from a user can be used to determine the cluster to which the user belongs to and accordingly utilize the pre-trained customized models and policies for the given cluster. This allows ease of deployment in commercial settings without the need to retrain the models for personalization.

\section{Conclusion}

We presented a trust dynamics-based clustering framework to identify and develop customized trust models based on dominant human trust behavior among a large population. We showed that such a framework can balance the tradeoff between a single general, or several personalized, models of human trust. We identified participants in the two clusters, namely ``skeptical'' and ``confident'' based on their trust behavior. We showed that customized models developed based on these clusters significantly outperforms a general \textit{one-fit-all} model in predicting human trust and take-over behavior during interaction with a driving automation. Furthermore, trust dynamics-based clustering approach is better than age-, gender-, or driving style-based approach in developing such customized model. Finally, we showed that the clustered participants' behaviors could be explained reasonably and may be coincident with established psychology of human trust. Future work could involve using these customized models for real-time trust calibration paradigms to improve human-automation interactions.

\section*{ACKNOWLEDGMENT}

We sincerely acknowledge Jain Research Lab and REID Lab at Purdue University for human subject study design and data collection.


\bibliographystyle{./IEEEtran}
\bibliography{ ./bib/IEEEexample}

\end{document}